# Archives and AI

## An Overview of Current Debates and Future Perspectives

Giovanni Colavizza

University of Amsterdam, Amsterdam, The Netherlands, g.colavizza@uva.nl

Tobias Blanke

University of Amsterdam, Amsterdam, The Netherlands, t.blanke@uva.nl

Charles Jeurgens

University of Amsterdam, Amsterdam, The Netherlands, K.J.P.F.M.Jeurgens@uva.nl

Julia Noordegraaf

University of Amsterdam, Amsterdam, The Netherlands, J.J.Noordegraaf@uva.nl

**ABSTRACT**

The digital transformation is turning archives, both old and new, into data. As a consequence, automation in the form of artificial intelligence techniques is increasingly applied both to scale traditional recordkeeping activities, and to experiment with novel ways to capture, organise and access records. We survey recent developments at the intersection of Artificial Intelligence and archival thinking and practice. Our overview of this growing body of literature is organised through the lenses of the Records Continuum model. We find four broad themes in the literature on archives and artificial intelligence: theoretical and professional considerations, the automation of recordkeeping processes, organising and accessing archives, and novel forms of digital archives. We conclude by underlining emerging trends and directions for future work, which include the application of recordkeeping principles to the very data and processes which power modern artificial intelligence, and a more structural – yet critically-aware – integration of artificial intelligence into archival systems and practice.

**KEYWORDS**

Archives, Recordkeeping, Machine Learning, Artificial Intelligence

## Introduction

Long before big data as an idea had been invented, archives already measured their collections in kilometers of files and folders. Worldwide large-scale digitisation efforts have by now transformed at least some of these collections into digital data. Next to these, from the 1990s onwards governments and other institutions with archival interests have increasingly worked digitally. This change did not immediately lead to a transformation of archival practice and workflows. The archival process still remained defined largely by manual appraisal, selection and review as long as the size of the collections and records still allowed this. However, the time window of this shift is closing fast. More and more archival collections are digitised and new born-digital records at ever larger scale are being submitted to archives. This makes a manual archival process less and less feasible. At the same time, records still need to be evaluated to ensure quality and trust, which provide the foundations of archives. Consequently, human archivists need the support of machine agents to assist them working through archival big data. The role of archivists is thus transformed, as they need to learn to make use of machine reasoning for appraisal and selection and to assess the assessments of machines. The archive becomes a big data organisation and like all big data organisations needs to at least partly put its trust into *Artificial Intelligence (AI)*, mostly in the form of machine learning, to deal with the transformation.

Across the world, archives acquire AI capacities to organise their workflows around the big data they have as well as to offer their big data to outside organisations. Ten years ago AI activities in archives were still largely experiments that showcased a potential, as they offered new ways of working with specific parts of the archival holdings like digitised newspaper collections. This has changed. We have recently observed a new trend where AI is used throughout the recordkeeping processes that characterise archives. *In this article, we survey recent research at the intersection of AI and archival processes, using the lenses of the Records Continuum model*.

To our knowledge, there is no such survey of the relationship between AI and archives. [Romein 2020] provides a detailed overview for the related field of digital history, while [Fiorucci 2020] contributes a closely related survey for cultural heritage in general. While neither survey considers the records' continuum that defines archives, they do observe that machine learning is adopted widely in the heritage sector in many spread-out experiments that target the specifics of individual collections, remarking how machine learning is now ubiquitous in memory institutions. [Binmakhashen 2020] is an example of a paper that surveys how machine learning can influence archival practices via the technical area of (automated) document processing, yet without taking a recordkeeping viewpoint. More closely related to our interests is [Marciano 2018]. The authors start from the same assumption that archival practice will be transformed by new advanced digital methodologies such as machine learning. They go through a range of case studies to describe a new interdiscipline at the intersection of archival and computer science and make detailed suggestions for changes to archival education.

Our contribution is organised as follows: we start by clarifying the scope of the survey, and our methodology to assemble related works. We discuss the literature, organising it in four broad thematic categories, which we identified. Finally, we critically discuss these trends and underline what future opportunities we see for this area of study.

## Scope of the survey

Writing a survey article on how in academic journals and literature the impact of artificial intelligence on (institutional) archives and broader (social) archiving processes is discussed, requires a clear demarcation of the field. We are not only interested in making a survey of the relationship between AI and archives located in the collections of heritage institutions, but in the relationship between AI and records in all its facets of creation, management and use. For that reason, we use the inclusive recordkeeping approach, which embraces the disciplines that have traditionally been distinguished in records management and archiving. Recordkeeping activities are targeted at creating, maintaining and accessing complete, accurate and reliable evidence of transactions which has the form of recorded information. Recordkeeping is grounded in core concepts such as provenance, context and evidence.

This comprehensive recordkeeping perspective is most clearly embraced by the *theory of the Records Continuum* [Gilliland 2014; Upward 2018]. When reviewing the literature discussing artificial intelligence in the field of recordkeeping, we are interested in how this translates into the *four dimensions of the Records Continuum*, as illustrated in Figure 1.

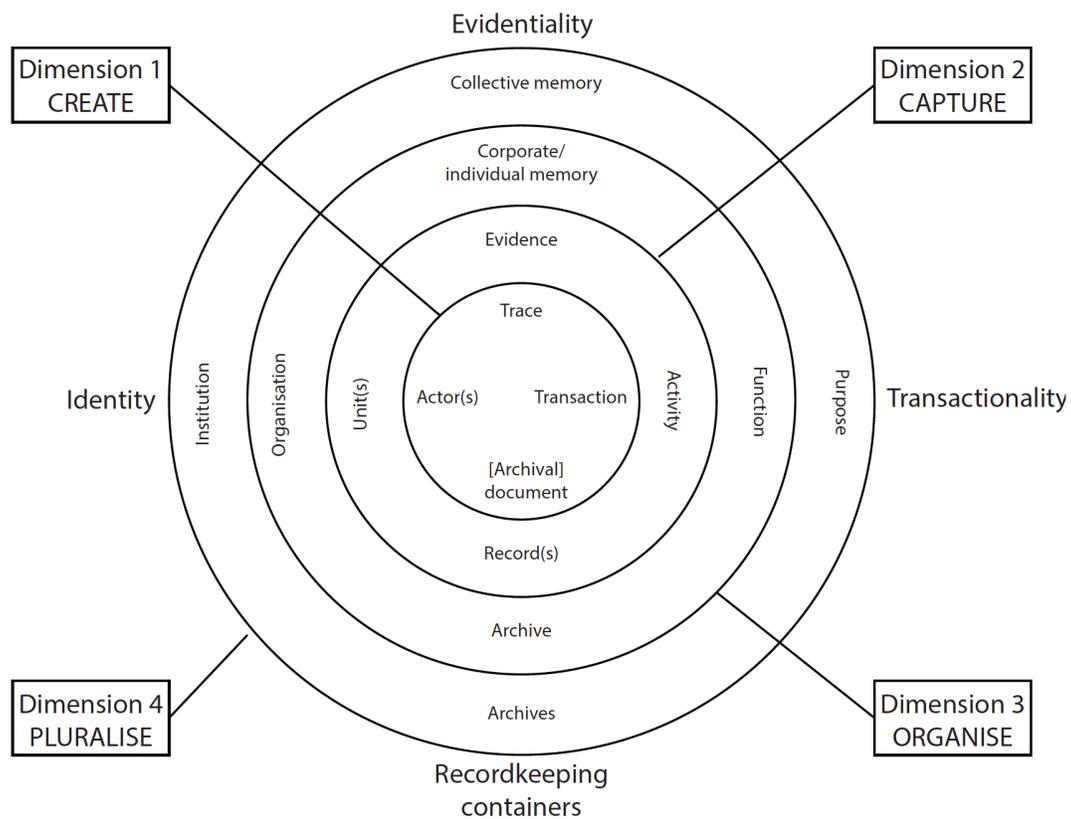

*Figure 1. The Records Continuum model [Upward 2018].*

The Records Continuum model provides a useful lens to structure the survey and to discuss the findings. The model offers a holistic, pluralist framework covering all aspects and dimensions of archiving: the creation of records, their capturing, organising and pluralising. The model not only includes records, but also actors and processes that underlie and affect the creation, management, organisation and use of records. Moreover, it presents these phases as circular and mutually influential. These dimensions have been a useful starting point for structuring the identified studies on artificial intelligence and archives because they allowed us, on the one hand, to identify exactly at which points of the archival process AI is employed and, on the other hand, to clarify how the use of AI influences the process of records creation, processing and reuse. Within our survey, the Records Continuum model was not only useful for categorising, but also for reflecting on the findings. After all, the Records Continuum is not a linear model that only relates to the lifecycle of records. The added value of the model is to understand how different dimensions may influence each other. Using records for new purposes may imply that records are re-created and take new meaning on the axes of evidentiality, identity, transactionality or recordkeeping containers. The Records Continuum model also helps to understand how activities and needs arising in one dimension may influence other dimensions; for example, how external societal expectations and norms (in the pluralise dimension) may have an impact on aspects of the creation, capture and organisation of records.

For this survey, we use Artificial Intelligence (AI) primarily as a proxy for Machine Learning (ML), that is the study and development of computer programs which automatically learn from data. While we acknowledge that more generally ML can be considered as an application of AI, we deem it to be the most impactful and directly related to recordkeeping, as previously defined. ML also relates to recordkeeping via other technical research areas, including Natural Language Processing (NLP), Computer Vision, Robotics, among others. Furthermore, we say artificial intelligence to make it clear that we are not only, or not even primarily, interested in technical systems, but instead in their application to recordkeeping – broadly construed – and what follows from it. We also use AI to encompass the professional, cultural and societal consequences of automated systems for recordkeeping processes and for archivists.

## Methodology

In order to systematically collect references for this work, we proceeded as follows. First, we compiled a list of publication venues – journals and conferences – for the research areas of archive and information science and digital humanities. This list is made available as an appendix (Table 1). While we acknowledge that many more venues and domains could fit this broad requirement – e.g., law, policing and surveillance, document processing – we deemed our selection to suffice as a starting point for representing the current debates on the topic at hand. For these venues, we sifted through all the issues of the past 6 years (2015 to 2020 included), up to the most recent one. Our motivation to select 2015 as a starting point has been two-fold: on the one hand, modern AI is largely driven by recent advances in so-called deep learning, or the use of neural networks. While work on neural networks dates back several decades, it first became mainstream when a system using convolutional neural networks brought a 41% improvement over the next (non-neural) competitor at the 2012 ImageNet Large Scale Visual Recognition Challenge (ILSVRC) [Krizhevsky 2012]. Given the time needed for these innovations to impact a relatively distant field such as recordkeeping, it seemed to us reasonable to start our survey in 2015. A cursory look at publications before this date sufficed to confirm our decision. Secondly, we wanted to focus on current debates and future perspectives, hence considering modern AI.

Our sifting focused on selecting contributions with a direct or otherwise explicit reference to recordkeeping theory and practice. Therefore, we did not include purely technical work on archival materials, in the absence of an archival reflection, nor archival work in the absence of an explicit reference to AI. This led us to individuate a core set of publications, which we further discussed and organised according to the Records Continuum model. Furthermore, we used these publications as a seed for reference lookup: we checked their list of references and the citations they received, in order to include more relevant materials even from outside of our original list of venues. Lastly, we complemented this expanded list with key and recent references which we deemed relevant according to our own expertise, and which come from outside the archive and information science or digital humanities literature.

The result is a selection of 53 references, which we analyse here. We intend this not as a systematic review, but as a representative overview of the topic at hand, and acknowledge the influence of our own perspectives into it.

## A Survey of Archives and AI

We organise the literature into four broad thematic sections: *Theoretical and professional considerations; Automating recordkeeping processes; Organising and accessing archives; Novel forms of digital archives*. This latter section also includes reference to recent emerging trends. Each section discussed related work using the Records Continuum model as framework of reference.

### Theoretical and professional considerations

An important question is how the proliferation of AI affects formative archival concepts and archival theory. This discussion is mainly conducted in the context of the overall digital transformation and less specifically from an AI perspective. [Moss 2018] consider the transformation of the archive from a collection of administrative records into a collection of data as one of the most fundamental consequences of the datafication. This reconceptualisation of the archive to data 'to be made sense of' means, among other things, that traditional methods of appraisal have become obsolete. Not only new human-machine-based tools are needed, but also new methods in the use and analysis of archives as data. [Upward 2018] announces the bankruptcy of managing records as end products and offers a new framework of recordkeeping informatics to support archiving processes. The reality of "cascading inscriptions" requires that information professionals engage with "nanosecond archiving in which everything is intertwined" and call for "new forms of professionalism that can authoritatively order the allocative power of new forms of computation" [Upward 2018].

There is a general awareness that the digital transformation has put pressure on archival concepts such as provenance and original order [Thibodeau 2016] and this generates debates about the relevance and meaning of ingrained "archival notions of the record, evidence, fixity, permanence, uniqueness, authenticity, ownership, and custody" [Gilliland 2014]. These discussions, Thibodeau argues, often lead to unproductive positions, while a different approach could be enriching. He shows how concepts in archival science can be enriched and reformulated by incorporating systemic functional linguistics (which provides a framework for understanding provenance through empirical analysis of context) and graph theory (capturing diverse relationships) and will improve understanding of records, including the creation and use. Another fundamental discussion revolves around the contribution of recordkeeping to transparency and accountability of records that are based on algorithmic output. This touches on the creation and capture of records. [Andresen 2020] offers a discussion frame to find ways how explanations to algorithms may be documented. The frame is based on concepts from records management, philosophy of science, and legal studies. In a similar vein, [Bunn 2020] uses the lens of eXplainable Artificial Intelligence (XAI), which requires to store not just the records but also the explanation. She argues that archives have to change – especially as the GDPR demands that the "meaningful information about the logic" involved in machine decision is made clear.

Looking at implications of AI for the profession, [Theimer 2018] sketches a picture of the future of the profession in which numerous archive tasks can and will be taken over by machines and algorithms. This implies that archivists need to become "masters of data" but at the same time the profession has to focus more on "narrative, storytelling, meaning-making, context providing (...) outreach and education [which] will be more important than ever" [Theimer 2018]. [Rolan 2019] observe an emerging gap between archivists who are familiar with AI concepts and those who are not and stress the necessity of "retraining" to mitigate the knowledge and skills gap. Lemieux advocates setting up a new 'transdiscipline' that draws archivists into the computational, and draws computer scientists and software engineers into the archival. This dialogue is reflected in an emerging field which has been labelled as Computational Archival Science [Lemieux 2018; Marciano 2018]. [Ranade 2018] argues that technical advances require a significant shift in archival thinking, which brings usability, trust and context in the center of archival work. Key components of this rethinking are acceptance of messiness and uncertainty and being transparent about the probabilistic nature and corresponding uncertainty of archival (meta)data. [Upward 2018] call for a new discipline base which requires new professional groupings, new skills and knowledge and conclude that "[t]he key to change lies with educators".

## Automating recordkeeping processes and decisions

One of the promising prospects of AI-technologies is the potential to automate parts of the archival workflow. Discussions and expectations which relate to this potential are predominantly situated in the 'Capture' and 'Pluralise' dimensions of the Records Continuum model. Although [Rolan 2019] envisage profound changes in archive work in the years to come, they are not very impressed with the progress in this field which can be attributed, among other things, to the difficulties of transcending one's own discipline. Designing accurate AI applications is based on data, domain expertise and data scientists who develop the algorithms. In addition, it should not be underestimated that significant groundwork (data preparation, developing workflows) needs to be done before AI can be successfully applied and this is illustrated with four Australian cases. Manual, high-quality and collaborative data curation remains a successful approach in some instances [Turchin 2015].

**Appraisal**

Much of the reviewed literature on automating archiving processes focuses on issues of appraisal. Although modern recordkeeping guidelines emphasise that appraisal is a proactive and not a reactive approach to the creation, capture and management of records, the reality is that recordkeeping professionals are confronted with ever expanding volumes of unstructured and non-categorised records. Artificial Intelligence has potential to assist managing such large volumes and several authors report on experiments and investigations. [Vellino 2016] describe experiments in two email corpora and demonstrate that automatic classifiers can successfully be trained to replicate expert's decisions for identifying e-mails with and without business value. [Lee 2018] observes some progress in applying software support in archival processes, but argues that much more could be achieved by deploying digital forensics, natural language processing and machine learning in developing better support in the vital recordkeeping domain of appraisal and he calls for more research. [Hutchinson 2020] provides this perspective by giving an overview of efforts of using NLP and machine learning in appraisal. He reviews several software tools (among others ePADD, BitCurator NLP, ArchExtract and a few commercial tools) and identifies and discusses five design principles (usable, interoperable, flexible, iterative, configurable) that NLP tools must adhere to in order to be successfully integrated into archival workflows. [Shabou 2020] report about an exploratory research in which a proof of concept and mock-up of an archival appraisal tool was developed to support the identification and extraction of structured and unstructured corporate data for timely disposal or transfer to an archival repository. The researchers designed a tool which combined a top-down archival approach (framework with appraisal dimensions, variables and metrics) and bottom-up data-mining approach (automatic topic classification, full indexing, Named Entity Recognition) to facilitate appraisal of large quantities of digital records and data.

**Handling sensitive information**

One of the most urgent challenges faced by government agencies and archival institutions is to prevent early disclosure of sensitive or personally identifiable information. At the same time, agencies want to avoid that an inability to identify sensitive information makes it virtually impossible to grant freedom of information (FOI) requests or to allow archival research. [Baron 2017] suggest various analytical approaches and techniques for identifying sensitive content and propose a research agenda for the archival community to develop technology assisted strategies. [Schneider 2019] describe how the software tool ePADD is used by five institutions for screening emails on sensitive, confidential or legally restricted information. Similarly, [Moss 2017] advocate developing assistance tools for digital sensitivity review, but they problematize the concept of sensitivity by arguing that it is far too simple to suggest that merely the subject of a document defines what is sensitive. Sensitivity is highly dependent on contexts in which documents are produced and used. Therefore, the authors call for research in which contextual analysis is incorporated, in order to be able to make nuanced decisions on sensitivity. Likewise, [Souza 2016] discuss the problems of identifying sensitive information and present an approach for automatic classification of documents in the archives of the state to check whether they can be released or should remain classified. [Hutchinson 2018] suggests that supervised machine learning could be a viable approach for a "triage" method of reviewing collections and identifying privacy sensitive records.

**Metadata**

Another area of recordkeeping processes in which AI is discussed focuses on issues of description and extracting metadata. [Büttner 2019] reports on a small, but successful proof of concept project in which semantic rule-based auto-classification for official documents in the Council of Europe was developed to relieve record creators from providing metadata, and increase the quality of subject metadata. [Spencer 2017] describes how automated methods can be applied to identify relationships between records and how these technologies may help provide additional contexts to records.

## Organising and accessing archives

The increasing amount of digitised and born digital archival materials call for the automation and even re-invention of the organisation and access to records. Several related contributions fall within the 'Organise' and 'Pluralise' Records Continuum dimensions, in the sense that they assume the existence of records which have already been created and captured. Yet, in so doing, they often create novel archival information in turn. A common theme is that AI can be used to move beyond the traditional principles of organising and accessing archives by provenance and original order, which focused on inventories and descriptions of archival units. This approach can now be complemented by other means, primarily focused on the contents of records. This shift not only allows users to make sense of increasingly larger archival materials, but also to do so in a variety of novel ways.

**Automatic content extraction and indexation**

The use of AI to re-think how archives can be organised starts with the automatic extraction of contents and their indexing. While our focus here is not on technical document processing – see [Binmakhashen 2020] for a recent overview – we single out the contribution of *Transkribus* in this respect [Muehlberger 2019]. Transkribus is a tool which integrates image and text recognition models to facilitate Handwritten and Optical Character Recognition. The tool, accessible via a graphical interface or programmatically through APIs, is designed to "fit neatly in the archival workflow, making direct use of growing repositories of digitised images of historical texts". Transkribus is increasingly being adopted by archives and other institutions, playing a crucial role in expanding the use of AI to extract records' contents and make them searchable. Access to records' contents, in particular texts, in turn allows using them for indexation. [Colavizza 2019] propose to automatically create archival information systems leveraging historical indexes, often produced when an archive was still current. Since these indexes often focus on entities such as persons, places or keywords, which occur across archival fonds, they effectively allow users to search an archive in a complementary way to provenance and original order. A similar intent is also the starting point of the *Traces through Time* project at the UK National Archives, which focuses on people as entities [Ranade 2016]. It was found that entity-based indexation, spanning across archival funds, allowed to search the archive in a more "fluid" way. Entity-based indexation is often served as linked data [Wilde 2017].

**Distant reading archival records**

A growing stream of work has focused on 'distant reading' [Moretti 2013] records' contents (see [Bode 2017] for a critique of the concept and related approaches in the context of literary studies). A popular method in this respect is topic modeling, which allows users to find 'topics' in large textual collections, defined by frequently co-occurring terms. [Chaney 2016] present the "capsule" model for analysing historical corpora containing events, with a case study on the US State Department cables. Their proposed methodology allows them to detect exceptional events deviating from common diplomatic parlance. Similarly, [Blanke 2017] use both topic modelling and language models (word embeddings) to extract and study "epochs" in textual archives. An epoch is defined by them as a period of coherent use of language. Epochs are first individuated using dictionary-based measures and then qualified and analysed using topic models and word embeddings. Another application of topic models is to extract keywords for indexation. [Hengchen 2016] apply this methodology to the European Commission archives. They extract keywords and then match them to an existing thesaurus, in order to enhance the discoverability of these holdings. Extracted keywords can also be enriched via language models, to further enhance the semantic indexation of archival contents [Coeckelbergs 2020]. Distant reading is prominent, given the primarily textual nature of archival records, yet other forms of contents are also being worked upon. For example, [Van Den Heuvel 2015] compellingly makes the case for indexing both texts and visual contents, such as drawings, which circulated through epistolary networks across the Republic of Letters. [Wevers 2019] offer an accessible overview of the possibilities of modern-day AI for 'distant viewing'.

**Search and retrieval**

Searching and retrieving information contained in archival records has also been an active area of work. [Lee 2019] uses visual template matching and automatic classification to retrieve reference cards from the International Tracing Service digital archive, one of the largest and most heterogeneous collections of Holocaust-related material. The need to aid archivists and other users to sift through possibly relevant records via automatic classification is also addressed by other predictive frameworks [Risi 2019]. [Bell 2020] takes, instead, a different approach based on archival catalogs. This work proposes to automatically aggregate archival description using text mining techniques, in order to aid their exploration. A complementary technique is used by the EHRI project (European Holocaust Research Infrastructure). The project automatically synchronises archival metadata from several sources, and complements them with user-generated metadata and thematic "virtual collections" [Bryant 2018]. Archival metadata is also used to organise archival materials in novel and richer ways; an example is the use of sentiment analysis applied to Germaine Greer's archive [Weber 2019]. Personalised retrieval is discussed by [Bocyte 2020] in the context of the Re:TV project, as an application of AI to support user engagement with audio-visual collections. Furthermore, AI methods are also being increasingly embedded into tools for the use of historians, allowing users to explore archival contents via information visualisation [Hinrichs 2015; Moretti 2016; Lansdall-Welfare 2020].

The application of AI for search and retrieval has been critically discussed by historians. Problematising the use of full-text search within born-digital archives, [Winters 2019] call for new approaches from archival science and artificial intelligence that are more sensitive to archival hierarchy and context, avoiding the pitfall of only finding what is known ('confirmation bias') in favour of seeking the unknown or surfacing gaps and absences in the data. In this sense, their contribution relates to other emergent reflections on born-digital archives [Chabin 2020] and the use of archival thinking in machine learning [Jo 2019], which we discuss below. A possible solution is "critical search" [Guldi 2018]. In her work, Guldi suggests that, rather than focusing on the application of a single tool to retrieve information from historical archives, a form of 'tool criticism' should be applied at every step of a historian's search process in digital archives. This process is characterised by a series of steps alternating applying algorithms, reading and critical reflection.

## Novel forms of digital archives

In addition to providing new opportunities to automate various dimensions of the archival process, the proliferation of digital technologies in all areas of society also generates new forms of digital archives that extend the 'Capture' and 'Organise' domain of archives as institutions into the 'Create' and 'Pluralise' dimensions of the Records Continuum model. Artificial Intelligence technologies play multiple roles in this process. First, they facilitate new types of data collection, as in the embedded, semi-intelligent systems of the Internet of Things (IoT). Based on two use cases (water-quality sensor data and automatic number plate recognition) [Sødring 2020] argue that government agencies using such IoT systems should make sure that the way the resulting information is captured and organised, complies with laws, regulations and expectations regarding privacy and other public values. The authors contend that the eight generally accepted record-keeping principles (accountability, transparency, integrity, protection, compliance, availability, retention and disposition) might provide guidance for organising these new forms of digital archive.

Similarly, [Chabin 2020] demonstrates how principles of archival science can help to improve the quality of data collection and processing outside legally defined record keeping contexts. Based on a case study, the body of documentation collected by the French government during the 'great national debate' that was launched as a response to the protests of the so-called 'yellow vests' movement in early 2019, Chabin shows how adopting a "diplomatic approach" to authenticating the documents could have made it possible to extract knowledge on the population of contributors and their behaviour. In addition, she argues that adding an unsupervised approach to the analysis of this born digital archive could have revealed "the diversity and richness of the citizen's voice in its entirety" [Chabin 2020], including topics and proposals not included in the reference frame of the government's official questions that initially guided the analysis. Here, AI technology and methods are presented as a means for realising a more democratic, inclusive form of data collection and analysis.

AI technologies also allow for the capture and organisation of records of events that would normally escape the legally or self defined collection policies of institutional archives. For example, [Sönmez 2016] describe a system for automatically crawling digital resources on social protests in Brazil, whereby a classifier detects protest-related reports in a newswire feed and entity extraction is used for indexing the records. The result is a new protest database which includes events such as strikes, rallies, boycotts, protests, and riots, as well as their attributes such as location, participants, and ideology.

In other cases, AI technologies and methods are propagated for analysing the new, vast collections of born digital records that emerge both within and outside established archival institutions. [Connelly 2020] discuss the use of NLP techniques for extracting metadata from diplomatic documents in the Freedom of Information Archive database. They demonstrate the potential of this approach by creating a new index for "country importance" in the context of US foreign policy priorities. [Milligan 2016] sees great potential in Web archives such as the Internet Archive for research on recent history because, in spite of the fact that websites are not representative of the entire population, they do record the voices of millions of non-elite individuals. [Black 2016] discusses the use of Web scraping as a means for analysing Web archival data, allowing scholars to contribute a humanities perspective to the study of the internet. He also contends that the lessons learned from Web scraping are useful for the text mining of large-scale digitised archival collections, such as newspaper collections. The Archives Unleashed Project [Ruest 2020] aims "to make petabytes of historical internet content accessible to scholars and others interested in researching the recent past" by providing access to thematic collections of websites archived with the Internet Archive's Archive-It tool and a toolkit to employ those resources in a research workflow. [Lynch 2017] shifts the perspective from the data and events to the tools used to capture and process them. He addresses the problem of stewardship for algorithms, and argues that current archival techniques and training are insufficient to preserve a perspective of the current 'age of algorithms' for the future. Lynch also provides a few preliminary suggestions on how to change the education of archivists in order to capture what algorithms do to us.

### Emerging trends

Attention to the role of AI in archival processes is not limited to the recordkeeping community. Recently, the topic has received attention from the perspective of AI technology development and implementation, in particular from scholars that reflect on the ethics of AI as a socio-technical practice. [Gupta 2020] explore the ramifications of centralised, due process archival systems on marginalised communities (consisting of people marginalised on the basis of culture, race, gender, age, physical or mental abilities). They propose to develop AI systems that are able to detect the low-traffic, fragmented pockets on the internet where some of the smaller communities built and maintain their own archives, as such pointing towards the use of AI for a more equitable, inclusive form of archiving. On the other hand, [Jo 2019] turn to the standards for archival selection and description that have been developed over centuries of recordkeeping and explore their potential as a model for curating less biased, more transparent and inclusive socio-cultural training data for machine learning. Finally, [Mohamed 2020] more generally explore decades of critical thinking in the Humanities and argue that this knowledge provides a conceptual framework for the development and implementation of a more inclusive, culturally responsive AI.

## Discussion

Our survey surfaced several trends of interest. According to [Moss 2018], the digital transformation is reconfiguring the archive from a collection of administrative records into a collection of data. Indeed, traditional archival principles and concepts are under pressure, with AI often used to probe their boundaries and even move beyond them. Notable examples are the principles of provenance and original order, traditionally used to organise and access records, which constitute but one approach to these ends when considering the vast digitised or born-digital archives we presently contend with. AI, in this sense, is used to bring alternative, more fluid orderings to the archive, for example by indexing it using the contents of the records. While archival principles discover their limitations, they also could find new life in a datafied world. Emerging challenges from the AI community call for the application of recordkeeping know-how to the data powering modern-day AI systems [Jo 2019]. Such issues include data provenance, appraisal, contextualisation, transparency and accountability: all topics where archivists have a rich tradition to offer. Taking these calls for action up will require re-thinking the training of professional archivists: from our survey it appears that, while this need is acknowledged, solutions are still wanting.

Looking at the use of AI as part of archival processes, we observe a rich and growing activity focused on automating them. AI is experimented with to automate, in part or full, key archival processes such as appraisal and metadata creation. While these efforts have already delivered practical results, or hold the promise to do so, we also note a still limited reflexive attitude towards AI. The limitations of AI in automating archival processes, not to mention its (sometimes unintended) consequences, are still largely to be discussed and addressed in the literature. A key example are the biases, and their ethical consequences, which might be introduced by AI decisions. Encouraging first steps in this respect [e.g., Bunn 2019], should not distract from the realisation that most discussion about the technical, ethical, and societal implications of using AI in archival processes are yet to take place.

A similar consideration applies for the use of AI to organise and access archives in novel ways. A rich set of contributions is bringing about a re-consideration of how to experience and use archives, and the information they contain. While a better integration of AI techniques in archival information systems remains to be developed, the work we have surveyed is encouraging in showcasing how AI can be applied to better serve the needs of a more varied user base. A direction for future work, which we find to still be underexplored, considers the implications of using AI to organise and access archives for researchers and scholars. In particular, there is ample room to design and develop AI-powered solutions to improve and enrich the way scholars can use archives. For this to happen, more interdisciplinary exchanges comprising archivists, AI researchers, and humanities scholars (e.g., historians) will need to take place [see Guldi 2018; Winters 2019]. Digital transformations and, in particular, the Web, are also broadening what is, or can be considered a record. Novel forms of born-digital archives, from the Internet of Things to social media, rely on AI not only to be organised and accessed, but even to be created and captured in the first place [e.g., Ruest 2020]. AI has also shown its potential for supporting more democratic and inclusive forms of archiving, for example by surfacing, preserving and making available records from underrepresented communities and minority groups [e.g., Gupta 2020]. Such novel forms of archives pose unprecedented challenges and opportunities to archivists, both in terms of re-thinking archival processes and sheer scale.

Through the lenses of the Records Continuum model, we have found that a large share of work within the scope of our survey focuses on the organise and pluralise dimensions, to a lesser degree on the capture dimension, and to a minimal degree on the create dimension. This is, to some extent, understandable: AI is primarily used to automate processes in existing archives, e.g., after their digitisation or when born-digital, and to improve their accessibility. Yet we note that, according to the Records Continuum model, records are in a constant "state of becoming": when using AI even at the outermost dimensions, archivists are creating new archival information in turn. The recognition of this 'continuum', not to mention its integration into archival processes and systems, is still largely to come.

In closing, this survey demonstrates how important interdisciplinary exchanges will be in the future. On the one hand, (digital) humanists should collaborate with archivists in order to create and deploy fit-for-purpose archival solutions leveraging AI techniques. On the other hand, computer scientists and archivists should jointly reflect on the implications of AI for recordkeeping and, vice-versa, on the contributions recordkeeping has to make to the advancement and appropriate use of data and AI, in and for society. In this respect, our survey leaves out key related areas of research – including AI and law and the ethics of AI – which might be profitably covered in future work.

## Conclusions

With archivists increasingly becoming custodians of larger and larger data sets and with no established approaches yet to address these new challenges, we have analysed the growing explorations of AI technologies for archives. This survey has provided a first and unique comprehensive overview of such work. We proceeded by seeking out approaches that have already had a significant influence on existing recordkeeping practices. Our contribution also lies in the methodology and structure we have given the work that has been surveyed, by framing it within the Records Continuum model.

Using the Records Continuum model, we have identified four distinct areas as well as future emerging trends. We started with the question how the integration of AI changes the theoretical foundations of archival practice. Here, we found several areas of overlap with more general concerns in archives about digital transformations. Beyond theoretical concerns, AI methodologies and technologies are already actively used to automate parts of the archival workflows, especially around the 'Capture' and 'Organise' dimensions of the Records Continuum model. Next, we investigated novel AI techniques used with regards to the 'Organise' and 'Pluralise' Records Continuum dimensions. There is much excitement within the archival communities about the potential of AI technologies here. In terms of organisation, AI helps archives break out of the limitations of traditional archival units while also maintaining their strengths, e.g., in terms of provenance. Future opportunities are to be found in terms of translating these strengths into the AI world. Archival provenance, for instance, needs specific techniques to be captured and maintained. AI techniques are also widely used to gain new insights from the now-digital collections, pluralise them through novel forms of retrieval and distant reading that allow archives to connect directly to new communities such as digital humanities. Finally, we described the role of AI technologies to open up new types of archives such as social media collections or diaries, which extend our current institution-focused ideas of archives. Because of their scale, these pose distinct challenges in terms of organisation and access to these collections that can only be solved with further AI technology development, which sets access to these collections apart from commercial Application Programming Interfaces (APIs). Only this way, the core archival mission of building trust in the past record can be sustained.

While much progress has been made, more research and direct work in and with archives is needed. We still see in particular gaps in terms of transforming the many experiments into lasting archival practice and infrastructure, and suggest that more work should concentrate on improving the trust into these AI techniques by developing a stronger ethical framework and a better understanding of their impact on research practices. We also think that more work is needed to update the theoretical foundations of archival practice with the recent developments in AI. Lastly, we single-out the opportunity for recordkeeping contributions to the advancement and appropriate use of AI, by bringing expertise on provenance, appraisal, contextualisation, transparency and accountability to the world of data.

## A    APPENDIX

*Table 1: List of venues. The column 'Latest issue' contains the latest volume, issue and year which we took into account for each venue, or the latest edition for conferences and workshops.*

| Venue | Type | Latest issue |
|---|---|---|
| AI & Society | Journal | 35, 1 (September), 2020 |
| Archival Science | Journal | 20, 3, 2020 |
| Archivaria | Journal | 89, Spring, 2020 |
| Archives and Manuscripts | Journal | 48, 2, 2020 |
| Archives and Records | Journal | 41, 1, 2020 |
| Artificial Intelligence and Digital Heritage: Challenges and Opportunities (ARTIDIGH) | Conference/Workshop | 2020 |
| Big Data & Society | Journal | 7, 2 (December), 2020 |
| Collections: A Journal for Museum and Archives Professionals | Journal | 14, 1, 2018 |
| COMMA | Journal | 2019, 1, 2020 |
| Computational Archival Science at IEEE Big Data | Conference/Workshop | 2018 |
| Computational Linguistics for Cultural Heritage, Social Sciences, Humanities and Literature ( LaTeCH-CLfL) | Conference/Workshop | 2019 |
| Conference on Natural Language Processing (KONVENS) | Conference | 2020 |
| DH Benelux Journal | Journal | v2, 2020 |
| Digital Humanities Quarterly (DHQ) | Journal | 14, 3, 2020 |
| Digital Medievalist | Journal | 10, 2015-16 |
| Digital Scholarship in the Humanities (DSH) | Journal | 32, 2 (June), 2020 |
| Digital Studies | Journal | 10, 1, 2020 |
| Document Analysis Systems | Conference/Workshop | 2020 |
| e-heritage workshop at ICCV | Conference/Workshop | 2019 |
| Fashion, Art, Design workshop recurring at ICCV/CVPR | Conference/Workshop | 2020 |
| Frontiers in Digital Humanities | Journal | 2020 |
| HistoInformatics | Conference/Workshop | 2019 |
| Humanist Studies & the Digital Age | Journal | 6, 1, 2019 |
| Information Discovery and Delivery | Journal | 48, 3, 2020 |
| International Conference on Frontiers of Handwriting Recognition (ICFHR) | Conference | 2018 |
| International Journal of Digital Curation | Journal | 15, 1, 2020 |
| International Journal of Digital Humanities | Journal | 2, 2019 |
| International Journal of Document Analysis and Recognition | Journal | 23, 3, 2020 |
| International Journal of Humanities and Arts Computing | Journal | 14, 1-2, 2020 |
| International Journal of Information Management | Journal | 52, June, 2020 |
| Journal of Archival Organization | Journal | 17, 1-2, 2020 |
| Journal of Cultural Analytics | Journal | July, 2020 |
| Journal of Data Mining and Digital Humanities | Journal | August, 2020 |
| Journal of Documentation | Journal | 76, 5, 2020 |
| Journal of Information Science | Journal | 46, 4, 2020 |
| Journal of Open Humanities Data | Journal | 6, 2020 |
| Journal of the Association for Information Science and Technology (JASIST) | Journal | 71, 10, 2020 |
| Journal on Computing and Cultural Heritage | Journal | 13, 3, 2020 |
| Proceedings of the FIAT/IFTA Media Management Seminars | Conference | 2017 |
| Records Management Journal | Journal | 30, 2, 2020 |
| The American Archivist | Journal | 82, 2, 2019 |
| The Journal of Contemporary Archival Studies | Journal | 7, 2020 |
| The Moving Image | Journal | 19, 1 (Spring), 2019 |
| Visarts | Conference/ Workshop | 2020 |
| Visual Pattern Extraction and Recognition for Cultural Heritage Understanding (VIPERC) | Conference/ Workshop | 2020 |